# Brain Functional Connectivity under Teleoperation Latency: a fNIRS Study

Yang Ye[1], Tianyu Zhou, Ph.D.[1], Qi Zhu, Ph.D.[2], William Vann, Ph.D.[1], Jing Du, Ph.D.[1, *]

[1] ICIC Lab, Engineering School of Sustainable Infrastructure & Environment, University of Florida, FL 32611, USA

[2] National Institute of Standards and Technology, Boulder, CO 80305, USA

[*] eric.du@essie.ufl.edu

Running head: Functional Connectivity under Latency

Manuscript type: Research article

Exact word count of text: 4805

**ACKNOWLEDGEMENTS**

This material is supported by the National Science Foundation (NSF) under grant 2024784 and The National Aeronautics and Space Administration (NASA) under grant 80NSSC21K0845. Any opinions, findings, conclusions, or recommendations expressed in this article are those of the authors and do not reflect the views of the NSF or NASA.

**Abstract**

**Objective**: This study aims to understand the cognitive impact of latency in teleoperation and the related mitigation methods, using functional Near-Infrared Spectroscopy (fNIRS) to analyze functional connectivity.

**Background**: Latency between command, execution, and feedback in teleoperation can impair performance and affect operators' mental state. The neural underpinnings of these effects are not well understood.

**Method**: A human subject experiment (n = 41) of a simulated remote robot manipulation task was performed. Three conditions were tested: no latency, with visual and haptic latency, with visual latency and no haptic latency. fNIRS and performance data were recorded and analyzed.

**Results**: The presence of latency in teleoperation significantly increased functional connectivity within and between prefrontal and motor cortexes. Maintaining visual latency while providing real-time haptic feedback reduced the average functional connectivity in all cortical networks and showed a significantly different connectivity ratio within prefrontal and motor cortical networks. The performance results showed the worst performance in the all-delayed condition and best performance in no latency condition, which echoes the neural activity patterns.

**Conclusion**: The study provides neurological evidence that latency in teleoperation increases cognitive load, anxiety, and challenges in motion planning and control. Real-time haptic feedback, however, positively influences neural pathways related to cognition, decision-making, and sensorimotor processes.

**Application**: This research can inform the design of ergonomic teleoperation systems that mitigate the effects of latency.

**Précis**: This study examines the cognitive impact of latency in teleoperation, using functional Near-Infrared Spectroscopy (fNIRS) to analyze functional connectivity. The study provides neurological evidence that latency in teleoperation increases cognitive load, anxiety, and challenges in motion planning and control. Real-time haptic feedback, however, positively influences neural pathways related to cognition, decision-making, and sensorimotor processes.

# 1. Introduction

Robotic teleoperation that involves a human operator controlling an intelligent system from a distance has gained an increasing popularity in difficult, dangerous and less accessible tasks (Neumeier et al., 2019). One of the critical needs in this area is to advance understanding of how latency affects human operators and the effectiveness of related mitigation methods. Latency in robot teleoperation is a prevalent, often inevitable issue and poses significant challenges to operators and system designers (Neumeier et al., 2019). The delay in data transmission between a human operator and a remote robotic system can result in reduced performance, decreased user satisfaction, and even safety concerns (Neumeier et al., 2019). This delay, known as latency, can stem from various sources, including hardware limitations and communication technologies. Mitigation approaches have been proposed to counteract the impact of latency when hardware and transmission technologies reach their limits, such as the predictive approach (Brudnak, 2016), adaptive control (Shahdi & Sirouspour, 2009), and multi-sensory feedback (Su et al., 2023). Predictive control strategies involve anticipating the system's future state to compensate for the delay (Brudnak, 2016; Sirouspour & Shahdi, 2006). Adaptive control techniques aim to adapt the robot's behavior in real-time to accommodate latency (Shahdi & Sirouspour, 2009), and multi-sensory feedback combines multiple sensory modalities to enhance the operator's perception and control capabilities (Su et al., 2023).

Among all approaches for mitigating the behavioral and cognitive consequences of teleoperation delays, the multi-sensory integration method in the control stands out as a promising solution. Unlike other methods that rely on automation and adaptive algorithms, multi-sensory control leverages sensory information to enhance the operator's subjective experience (Toet et al., 2020; Triantafyllidis et al., 2020), thus facilitating human adaptation to time-delayed teleoperation. Recent advancements in multi-sensory control include the integration of modified haptic feedback (Du et al., 2023), which plays a crucial role in mitigating the challenges associated with latency, such as robustness, reliability, and trust issues. In one of our previous studies (Du et al., 2023), we designed and tested a system that generated synthetic, simulated haptic feedback, coupled, or decoupled with the visual feedback in a teleoperation task for NASA's R&R (replacement and repair) task. Our preliminary analysis found that providing synthetic haptic feedback (e.g., momentum of moving an object in the direction of acceleration) immediately after the motor actions in real time could bring a variety of performance and perceptual benefits, such as increased accuracy of object placement, reduced time on tasks, and shorter perceived delays (Du et al., 2023). Although these findings provide promising evidence for identifying a novel mitigation method for improving the performance in delayed teleoperation, it is less clear how these multi-sensory variations impact operator's neural functions as a possible cause of changed motor and cognitive performance. Especially, the neural basis of how operators integrate synthetic haptic feedback with other sensory modalities at the neurofunctional level remains underexplored. Understanding the neural mechanisms that underlie this integration is crucial for designing more effective user interfaces. This would not only enhance immediate task performance but also ensure sustainable cognitive adaptation in other teleoperation scenarios. There is a need for exploring the changes in neural activity and cognitive processing associated with multi-sensory feedback in teleoperation tasks.

This study aims to bridge the research gap between the increasing occurrence of latency in teleoperation and the limited understanding of its cognitive impact and related mitigation methods. To achieve this, we conducted a simulated robot arm pick-and-place task within a physics engine, introducing data transmission latency. Participants were tasked with operating the robot through a haptic device while observing camera views from a control deck. Three controlled conditions were tested: a no-latency condition (NL), a delayed condition with synchronized latency (Syn), and an anchoring condition (Anc) (Du et al., 2023). NL condition simulated an optimal condition that transmitted the haptic and visual feedback with no latency. Syn condition represented a synchronized data transmission strategy in which all data were transmitted in the same batch with the same latency. Anc simulated an asynchronized data transmission with simulation strategy (Du et al., 2023), which simulates low-rank haptic feedback in local workstations, thus providing real-time haptic feedback. Anc condition was essentially a latency mitigation strategy in which we manipulated the users' sensory feelings to create a subjective feeling of less delay.

To investigate the cognitive impact of latency and the effectiveness of mitigation methods, this study utilizes Functional Near-Infrared Spectroscopy (fNIRS) for analyzing brain activities and underlying cognitive behaviors (Tak & Ye, 2014). Compared to medical neuroimaging techniques relying on neurovascular coupling, such as functional magnetic resonance imaging (fMRI) (Logothetis, 2008; Ochsner et al., 2002) and positron emission tomography (PET) (Andreasen et al., 1996), fNIRS demonstrates better portability and flexibility in field research. In addition, unlike sensors based on the electromagnetic activity of the brain, such as electroencephalography (EEG) (Ray & Cole, 1985) and magnetoencephalography (or MEG) (Halgren et al., 2000), fNIRS is more robust for motor artifacts (Vitorio et al., 2017) because it tracks cortical hemodynamics that is less sensitivity to body movements (Pinti et al., 2018). These features of fNIRS make it suitable for monitoring the operator's neural activities. fNIRS measures hemodynamic responses, including oxy-hemoglobin (HbO), deoxy-hemoglobin (Hb), and total hemoglobin (THb), providing insights into neural processing. Previous studies have demonstrated the utility of fNIRS in brain functional connectivity research (Nguyen et al., 2018; Tyagi et al., 2023), revealing networks within the sensorimotor, visual, auditory, and language systems during various tasks. Analyzing brain connectivity through fNIRS offers reliable and valuable insights into cognitive processes. A human-subject experiment involving 41 participants was conducted to collect performance data and fNIRS measurements. Functional connectivity analysis was performed to gain insights into neural activities within and between brain regions in response to latency and mitigation strategies. This research aims to provide valuable contributions to the field of teleoperation by shedding light on the cognitive aspects of latency and offering insights into effective mitigation techniques.

## 2. Method

### 2.1 Participants

We recruited 41 healthy subjects (college students) to participate in this experiment. Among all participants, there were 26 males (63.41%) and 15 females (36.59%). The average age was 25.12 ($\sigma$=9.34). A total of 18 were with engineering backgrounds and majored in civil engineering or mechanical engineering, while others were self-identified as non-engineering students. In addition, 12 participants (29.27%) self-identified with previous experience with VR, while others claimed a lack of experience with VR. To be noted, our post-experiment analysis did not find any difference among these demographic or experience groups. This research complied with the American Psychological Association Code of Ethics and was approved by the Institutional Review Board at University of Florida. Informed consent was obtained from each participant.

### 2.2 Experiment environment

To perform a controlled experiment, we built the experiment task in virtual reality (VR) in the Unity game engine (Zhou, Xia, et al., 2023). As shown in **Fig 1**, we simulated a teleoperation station in VR. Three simulated monitors were placed on the wall, providing front, side, and top-down views. Participants sat in front of the screens and manipulated a haptic controller, TouchX (3DSystems, 2022). The position and orientation of TouchX's end effector were mapped to the remote robot's end effector. Meanwhile, the pressure that participants exerted on TouchX's stylus was mapped to the remote robot's gripping force. The study was performed in a quiet room with constant light conditions. With VR and fNIRS sensors on, participants sat stably on a chair in front of a table and grabbed TouchX's stylus steadily. The design details of hardware and software systems, including the coupling of VR and haptic devices and robot arms, robot contact simulation, and inverse kinematics calculation, can be found in our previous papers (Ye, You, et al., 2023; Ye, Zhou, et al., 2023; Zhou, Xia, et al., 2023; Zhou, Zhu, et al., 2023) .

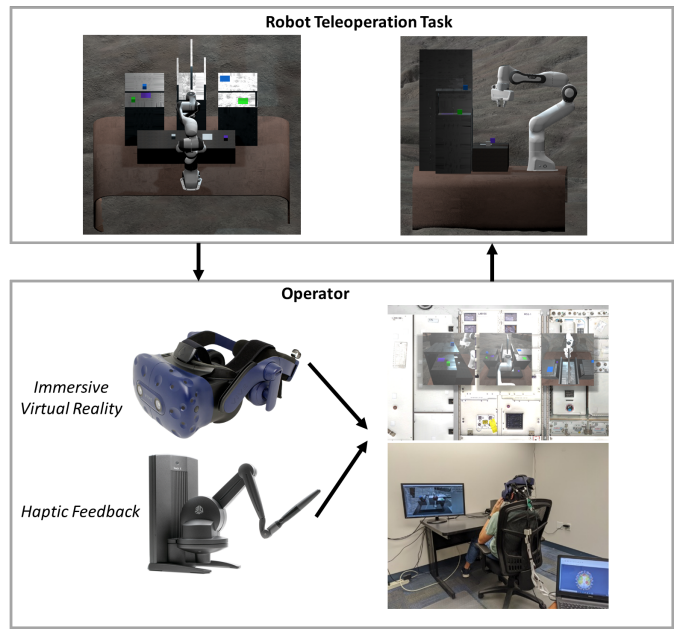

**Fig.1** The experiment environment for this research

## 2.3 Task design

The teleoperation task's objective was to grab and place objects in the correct position. There were four cubes with distinctive colors, and each cube was paired with a target position highlighted by the same color. The purpose was to show clearly where each cube should be moved. Participants had to pick up the cubes and place them on the corresponding targets in a fixed order for a controlled experiment. **Fig.2** illustrates the layout of the experiment task. The distances between each cube and its target varied, with obstacles strategically placed along the movement trajectories. These obstacles, differing in size and position, added complexity to the task, representing varying movement challenges that participants had to navigate.

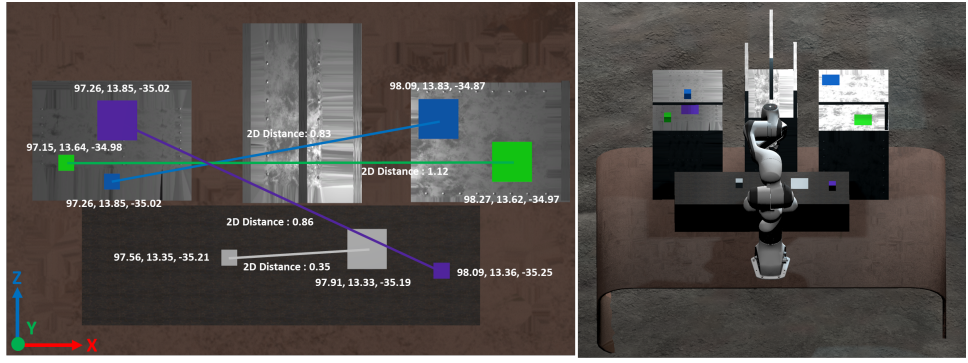

**Fig.2** The layout of the object manipulation task in the experiment

### 2.4 Latency design and experiment conditions

For a single discrete event, there are three latencies in this experiment. First, the latency between an action initiation by the participant and the execution by the robot ($\Delta_o$). This latency accounts for the command transmission lag from participants to the remote robot, and the time required for running inverse kinematics to calculate variable robot joint parameters. Second, the latency between the robot's haptic status ($\Delta_h$), such as touching an obstacle, and the haptic feedback received by participants. Similarly, there is a latency between capturing the visual information around the robot and the participant receiving the visual information in the workplace ($\Delta_v$). A time-series combination of these time points forms the event design with latency ($S$):

$$S = (t_a, t_r, t_h, t_v)^T = (t_a, t_a + \Delta_o, t_a + \Delta_o + \Delta_h, t_a + \Delta_o + \Delta_v)^T \quad \ldots \text{Eq (1)}$$

, where $t_a, t_r, t_h, t_v$ are the continuous time in which actions is initiated, executed, triggered haptic feedback, and triggered visual feedback. Different combinations of latency values form different conditions in this experiment. We take relevant conditions from the experiment, specifically:

NL condition assumes neglectable transmission and processing latency. In this condition, haptic and visual feedback both happen immediately after an action initiated by the human operator, i.e., in real-time. This simulates an optimal teleoperation scenario without the impact of latency. The latency design matrix is defined as:

$$S_{NL} = \lim_{\Delta \to 0}(t_a, t_a + \Delta, t_a + 2\Delta, t_a + 2\Delta)^T \quad \ldots \text{Eq (2)}$$

Anc condition assumes no delays for haptic feedback, and non-neglectable latency for visual feedback. This condition transmits haptic stimulation as soon as possible via simulated synthetic haptic cues. Due to data size, transmitting and decoding visual feedback takes longer. In this experiment, we take $\Delta_v$ ranges from 100 to 850 ms. We intend to mitigate the subjective feeling of latency through no-latency haptic feedback. This condition can be described as:

$$S_{Anc} = \lim_{\Delta_o \to 0}(t_a, t_a + \Delta_o, t_a + \Delta_o, t_a + \Delta_o + \Delta_v)^T \qquad \text{... Eq (3)}$$

Syn condition assumes non-neglectable latency and the same for both visual and haptic feedback. In this experiment, we take $\Delta_o = 150$ ms, $\Delta$ ranges from 100 ms to 850 ms. In this condition, the haptic feedback is intentionally delayed, matching the delayed visual feedback and creating synchronized feedback. The rationale is to ensure multisensory congruency, i.e., a coherent representation of sensory modalities to enable meaningful perceptual experiences. This condition can be expressed as:

$$S_{Syn} = (t_a, t_a + \Delta_o, t_a + \Delta_o + \Delta, t_a + \Delta_o + \Delta)^T \qquad \text{... Eq(4)}$$

The experiment was designed as a within-participant experiment, i.e., each participating subject experienced three conditions. To avoid learning effects, the sequence order was shuffled for each subject. Participants were told that latency issues existed in the experiment, but they did not know the latency configurations in each trial. The performance data (time and accuracy), motion data (moving trajectory), eye tracking data (gaze focus and pupillary size), and neurofunctional data (measured by fNIRS) were collected. Participating subjects were also requested to report their perceived delays to compare them with actual ones. Before the experiment, each participant was required to fill out a demographic survey and the consent form approved by UF's IRB office. Then, they would take a 10-minute training session to familiarize themselves with VR. Afterward, participants were required to take a break of 5 minutes by sitting quietly with all sensors on. This break session was for collecting fNIRS baseline data and removing possible impacts of the training session.

## 2.5 Performance data analysis

We first analyzed the performance of the object manipulation, including time on task measured in seconds and placement error measured in cm. Time on task is the difference between the end and start times, measuring the completion speed. The placement error is calculated as the Euclidean distance between the cube's actual placement and the target location's center. The smaller this distance, the more accurate the placement. Without further notes, all analyses are based on the aggregated data of four cubes.

Then, we analyzed the accuracy of subjective delay estimation. Two metrics were analyzed: visual perception difference and haptic perception difference. Visual perception difference is defined by the difference between the perceived and actual visual delay. Haptic perception difference is determined by the difference between the perceived and actual haptic delay. These evaluation metrics provide a comprehensive evaluation of the objective performance and subjective feeling of latencies.

## 2.6 fNIRS Setup and Analysis

fNIRS has proved to be an effective and scalable tool that measures the changes in oxygenated and deoxygenated hemoglobin concentration of the brain's cortical regions (Wilcox & Biondi, 2015) at a similar level to fMRI (Nguyen et al., 2018). Previous studies have demonstrated the significant correlation between functional connectivity measured by fNIRS among cortical areas and the level of depression and anxiety (Zhang et al., 2022), cognitive load (Newton et al., 2011), creativity (Wei et al., 2014), and certain disease severity (Wang et al., 2014). In this study, we are particularly interested in the impact of teleoperation latencies on cognitive and motor coordination functions. Previous studies suggest that the prefrontal cortex is the cognitive center of the brain, and the activity within the region is an indicator of mental workload and levels of anxiety and depression (Zhang et al., 2022). The motor cortex is believed to be the major cortical region facilitating motor control and coordination (Koch et al., 2006; Machado et al., 2010), which is the main contributor to generating neural impulses that pass down to the spinal cord and control movement execution (Rizzolatti et al., 2002). As such, we selected the prefrontal cortex and motor cortex as the central regions of interest.

The fNIRS head cap (Nirx, 2023) used in this study had 16 sources and 15 infrared light detectors, forming 40 channels. An additional detector was placed at the right pre-auricular point (RPA) as a reference point. The probe map is shown in **Fig 3**, visualized by MNE. The fNIRS setup covers the prefrontal cortex and motor cortex. Specifically, fNIRS in this study covered seven regions within the prefrontal and motor cortex defined in Brodmann areas: anterior prefrontal cortex (APFC), left dorsolateral prefrontal cortex (LPFC), right dorsolateral prefrontal cortex (RPFC), left premotor cortex (LM1), right premotor cortex (RM1), left primary motor cortex (LPM), and right primary motor cortex (RPM). APFC, LPFC, and RPFC were merged to evaluate prefrontal cortical activities. LM1, RM1, LPM, and RPM were merged to assess the motor cortex activities.

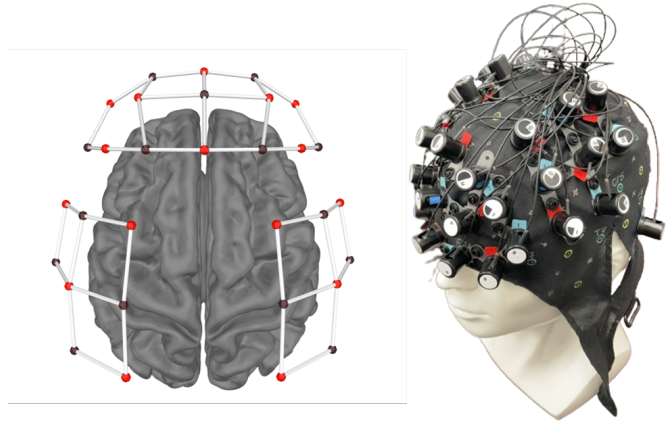

**Fig. 3**: fNIRS setup in this experiment

fNIRS data was preprocessed with MNE (Gramfort et al., 2013). Figure 4 shows the analysis pipeline. The time-series fNIRS data was read into the software and converted into optical density. Next, we measured the connection quality between the optode and the scalp through the scalp coupling index (SCI) (Pollonini et al., 2014). Channels in experiment trials with SCI less than 0.5 were marked as channels with low data quality and removed. The corrected signal was then bandpass filtered for a frequency range of 0.04 Hz–0.15 Hz [35] to remove physiological noise, with a transition bandwidth of 0.1 Hz and 0.02 Hz at the high and low cut-off frequency. Then, the filtered optical density signal was converted to the oxygenated and deoxygenated hemoglobin concentration according to the Beer-Lambert law with a partial path length factor of 6. Although fNIRS data can be analyzed with multiple measures such as oxy-hemoglobin (HbO), deoxy-hemoglobin (Hb), and total hemoglobin (THb), HbO was selected for further analysis because it is the most sensitive indicator to cerebral blood flow changes in motor-related tasks (Hoshi et al., 2001). We utilized the signal space projection (SSP) to suppress interference and noise in electromagnetic signals. After completing the preprocessing for fNIRS, we performed baseline correction using the resting-state data collected at the beginning of the experiment. In our analysis we focused on the pick-up evens (for continuous 40s worth of data prior and post picking up of the cubes), given its motor difficulties in the entire teleoperation task presented with delays. The pick-up events were marked in fNIRS data during the experiment. Each continuous trial was sliced into 40-second epochs, starting at 10 seconds before the events and ending at 30 seconds after the events.

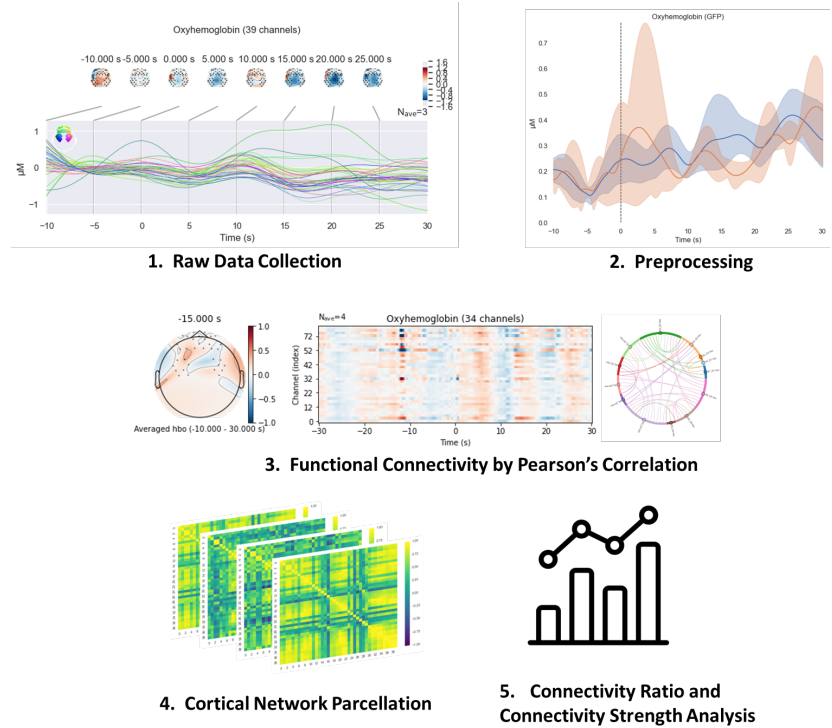

Figure 4: fNIRS Neuroconnectivity Analytical Pipeline

The Pearson correlation coefficients (ρ) were calculated between each pair of fNIRS channels for each epoch. The ρ values for a subject in each condition were the average of all ρ values calculated from the epochs belonging to this subject and this condition. In this experiment, the data generated a 40 × 40 ρ matrices for each epoch. Then, the ρ values were converted to a normal distribution through the Fisher z-transformation using the equation below:

$$z = \frac{1}{2}\ln\frac{1+\rho}{1-\rho} \quad \ldots\text{Eq (5)}$$

To deliver a more comprehensive evaluation, a cortical network parcellation was performed to divide the channels into two sub-networks in the Brodmann area: the prefrontal cortex (PFC, 20 channels) and the motor cortex (M1, 20 channels). Finally, four cortical networks were formed: 3 within-networks (full connection, PFC, and M1) and one inter-network (PFC-M1). The full connection network is 40 × 40 matrices, and the other networks (PFC, M1, PFC-M1) are 20 × 20 matrices.

Two features were extracted from the network metrics to perform the statistical analysis: connection ratio (CR) and connection strength (CH) (Nguyen et al., 2018). CR was calculated as the ratio of significant connections over all connections in a network. The significant connection was defined as the pair of channels with an absolute z-value greater than a threshold of 0.3 in this experiment. CS was defined as the average strength of connections in a network, calculated by the sum of absolute z values divided by the number of connections.

The number of significant connections in a network was calculated by counting all connections belonging to the network that had an absolute z-value greater than the threshold. In this study, the threshold range was chosen from 0.2 to 0.7. The connection ratio (CR) was then computed as a ratio of the significant connections on the network total connections. A network's connection strength (CS) was the average of the absolute z-values of all connections.

### 2.7 Statistical Analysis

The variables were individually first checked for normality, and those whose distributions differed from normality were transformed to a normal distribution using logarithmic transformation. Normally distributed variables were checked for homogeneous variances. Variables that passed the assumption checks were analyzed using analysis of variance (ANOVA) on parametric study measures. T-tests with Bonferroni corrections were used for post hoc pairwise comparison for each condition pair. Signed rank Wilcoxon tests were performed to examine any significant differences for non-parametric data distributions, including CR and CS. The statistical tests were considered statistically significant when the p-value was less than 0.05.

## 3. Results

### 3.1 Performance and Perceived Latency

**Fig.5** shows the result of performance and perceived latency. The results indicate significant differences in placement accuracy between the NL and the Syn conditions (p=0.004) and between the Anc and Syn conditions (p=0.032). We found that the Anc condition, i.e., providing real-time haptic cues, significantly improved the hand-picking task regarding placement accuracy. The benefits could be because participants could rely more on haptic feedback when it was available to coordinate the teleoperation actions. The advantage of providing real-time haptic stimulation boosted the performance to a level similar to the NL condition (p=0.168). As for time on task, the results also indicate significant differences between the NL condition and the Syn condition (p<0.001) and between the Anc condition and the Syn condition (p=0.049). While the Anc condition didn't perform as well as the NL condition (p=0.009), it still outperformed Syn conditions regarding time on task.

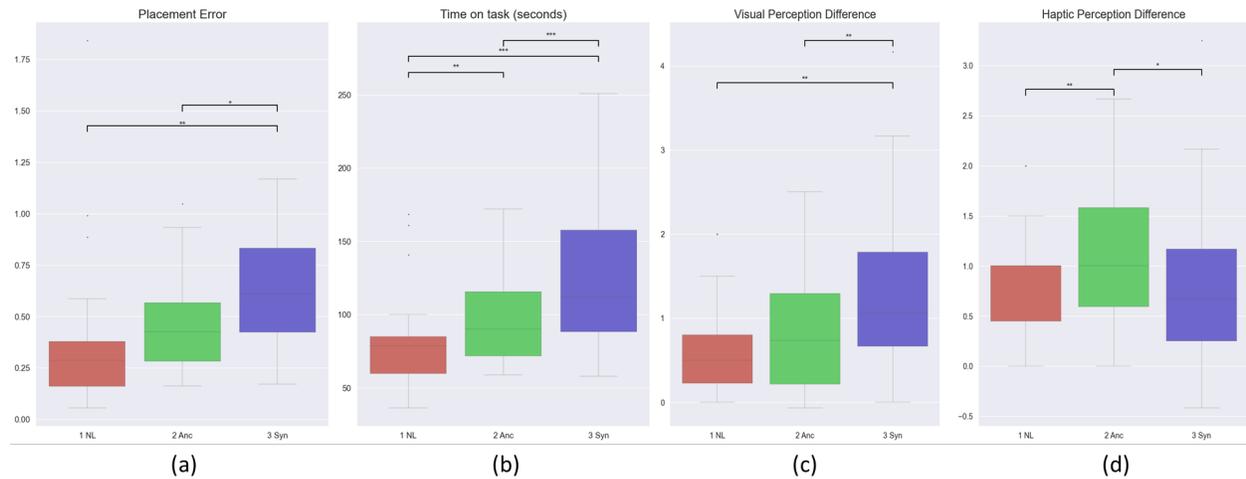

**Fig.5** Teleoperation performance measured by (a) placement error, (b) time on task comparison; Perceived latency accuracy: (c) visual perception difference, (d) haptic perception difference. *, **, and *** denote a pairwise comparison with 0.05 > p > 0.005, 0.005 > p > 0.0005, and 0.0005 > p

As for the perceived latency, the results suggest that the Anc method could reduce the subjective feeling of teleoperation latencies (visual delays) by up to 1 second. Here we focused on examining if the proposed sensory manipulation method could reduce perceived visual delay, as it is considered the most common and troublesome delay in robot teleoperation. The data shows that the overall average perceived visual delay in teleoperation under the Anc condition was significantly lower than the Syn condition. In addition, 18% of participants reported a perceived visual delay that was minor compared to the actual one under the Anc condition. It implies that coupling real-time haptic feedback with the action during teleoperation can mitigate the subjective feeling of delays. For example, when the actual visual delay was 750ms, a subject reported 100ms as the perceived delay. **Fig. 5c** and **Fig.5d** compare the haptic, visual perception difference, and the visuomotor gap perception difference. As for the perceived haptic delays, the data shows a slightly different pattern. Subjects seemed to report a lower perceived haptic delay under the Syn condition. This makes sense because the coupled haptic and visual feedback may help to estimate the haptic delay better. As for the visuomotor gap perception shows that under the Anc condition, many subjects reported a delay more minor than the actual one. All these results confirmed the perceptual benefits of having real-time haptic feedback.

## 3.2 Neural Connectivity

### 3.2.1 Connectivity Ratio

Figure 7 displays the connection ratios (CR) derived from HbO for four networks in three conditions. The CR threshold of 0.3 was chosen because previous studies (Nguyen et al., 2018) showed that 0.3 preserved a reasonable detail of connectivity patterns. In general, the presence of latency increased the CRs, as demonstrated by the comparison between the NL and Syn conditions. The Syn condition led to the highest CR, and the NL condition led to the lowest CR in all four networks. The CRs in Anc condition were in between. The ANOVA and post hoc tests verified the significant difference.

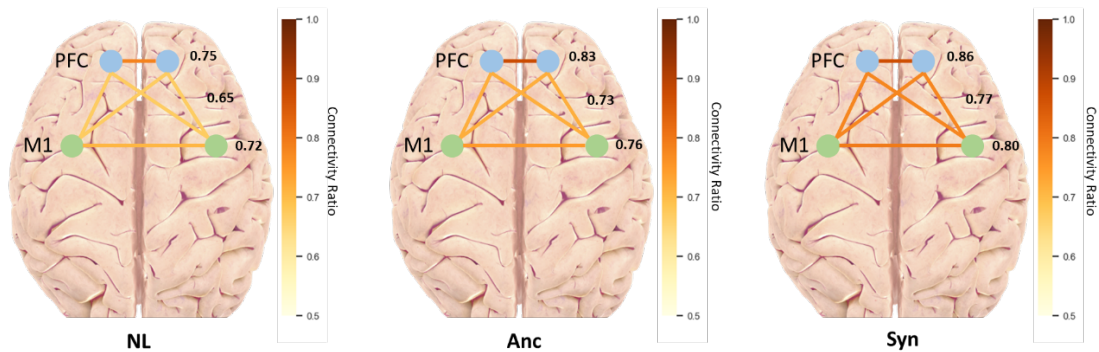

**Fig. 6** Average fNIRS CR values

Specifically, the participant's CR in the optimal teleoperation condition, NL, was significantly different with Anc and Syn in all networks of interest: full connection (p = 0.001; p <0.0005 compared with Anc and Syn), PFC (p = 0.002; p < 0.0005), M1 (p = 0.021; p = 0.001), and PFC-M1 (p = 0.002; p < 0.0005). Meanwhile, providing haptic feedback of lower latency led to significantly lower CRs compared with the synchronized haptic and visual latency within the PFC (p = 0.028) and M1 (p = 0.035) networks. Although the CRs of the full connection network (p = 0.073) and between PFC and M1 () were not significantly different between Anc and Syn conditions, the p-values were comparatively small (p = 0.073 for the full connection network, p = 0.238 for PFC-M1 network) and the average CRs were lower in Anc condition.

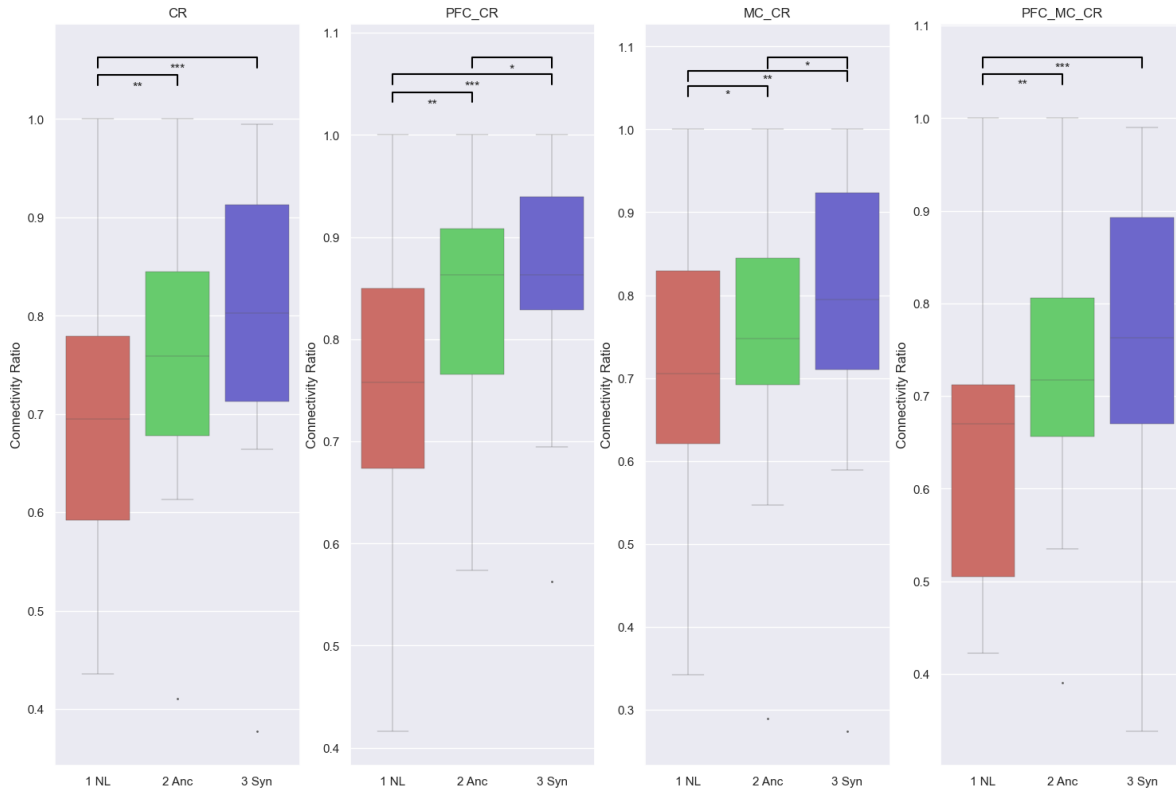

**Fig.6** fNIRS CR from HbO, with CR threshold = 0.3. *, **, and *** denote a pairwise comparison with 0.05 > p > 0.005, 0.005 > p > 0.0005, and 0.0005 > p

### 3.2.2 Connectivity Strengths

Average connectivity strengths calculated from HbO were generally similar, except between the NL and Syn conditions in the full connection network (p = 0.0076) and PFC network (p = 0.0036). Like the CR results, the average connectivity strengths were the lowest. At the same time, no latency was presented (NL condition) and highest while the same high latency for haptic feedback and visual feedback was shown (Syn condition).

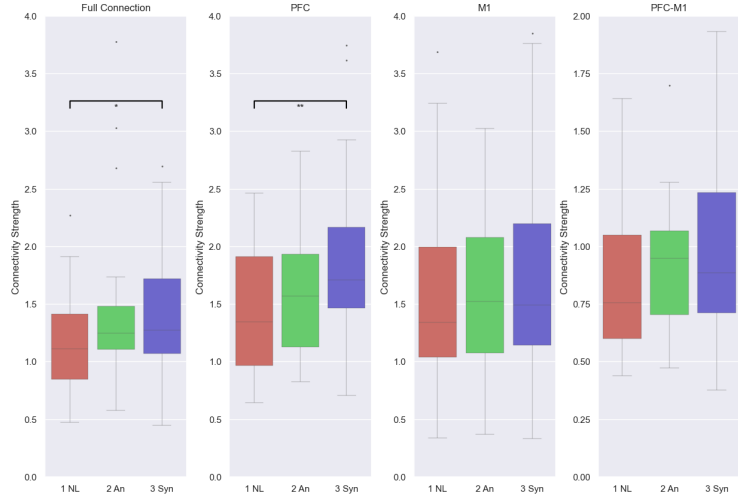

**Fig.7** fNIRS CS from HbO. *, **, and *** denote a pairwise comparison with $0.05 > p > 0.005$, $0.005 > p > 0.0005$, and $0.0005 > p$

## 4. Discussions

We designed an experiment to gain neurological evidence on the impact of teleoperation latency on human cognitive activities. The brain functionality networks were investigated under different teleoperation latencies: no latency (NL), real-time haptic feedback and full-delayed visual feedback (Anc), and haptic and visual feedback with synchronized latency (Syn). ANOVA and post hoc tests showed that the brain functional connectivity was strong in the Syn condition and weaker in the NL condition among all brain regions of interest (overall connection, within PFC, within M1, and between PFC and M1). This result provides solid neurological evidence that latency can significantly impact the neurological path and coordination among brain regions.

In particular, a statistically higher CR was expressed in all networks, and a significantly stronger CS was shown in the full connection and PFC networks in the Syn state. The increased functional connectivity within the PFC indicates a potential increase in cognitive load and anxiety level (Zhang et al., 2022), which echoes previous operation latency studies (Yang & Dorneich, 2017). The high functional connectivity between PFC and M1 implies the more intensive involvement of neuropaths that coordinate the perception and motor control functions, which can be interpreted as the increased difficulty of motor control and decision-making in this motor-intensive teleoperation task (Koch et al., 2006). Meanwhile, functional connectivity within M1 increased as the latency was introduced. This behavior implies more intensive involvement of motion selection and preparation [30] when latency is introduced. Taking the results together, this experiment showed neurological evidence that latency in teleoperation could induce cognitive burden, anxiety, and more challenges in high-level and motor-control decision-making. These mental activities unrevealed by fNIRS data coincide with the performance results that show that the presence of latency reduces teleoperation accuracy and speed.

Moreover, we evaluated the cognitive impact of a latency mitigation method that provides real-time haptic feedback to simulate a pseudo-real-time haptic perception for a more coordinated action-haptic perception loop (Du et al., 2023). This latency mitigation method aims to manipulate the operator's subjective perception of latency and achieve better performance and subjective feelings. The Anc condition in this experiment adopted such a mitigation principle. In general, results showed that the Anc condition reduced the CR and CS compared to Syn, attributing to the reduced average values and distribution statistics. In other words, Anc condition pushed the neurological patterns to a state closer to the no-latency teleoperation. Together with the performance improvement, the results provide neurological evidence that the mitigation methods can positively influence cognitive activities in the teleoperation task.

Interestingly, when latency is presented in teleoperation, CRs within PFC and within M1 were significantly reduced by the pseudo-real-time haptic feedback (Anc condition) but not between PFC and M1. These functional connectivity changes implied that neuropathy connecting different areas within PFC and within M1 cortexes were less activated, while the communications in between remained relatively intensive. A potential interpretation is that while maintaining a similar level of information exchange between decision and action, the Anc condition led to a paradigm shift in the neural process, in which operators relied more on the sensorimotor process, thus reducing the cognitive activity and motion decision load (Bardouille & Boe, 2012). This interpretation echoes the observation that participants reported less perceived visual delay in the Anc condition.

Limitations of this study need to be addressed to get a clearer understanding of the scope of generalizability of the results and to give direction to similar investigations in the future. First, this study presented evidence on the neural mechanism of latency-induced performance in teleoperation tasks. While these findings are essential, further studies are needed to investigate whether similar neural change patterns are valid for more complex teleoperation tasks and latency-mitigation methods. These mechanistic investigations are critical for designing future HMIs that understand and mitigate latency issues. Second, the experiment in this study was conducted as a single 2-hour session for each participant, which did not capture the learning effect and residual effect. Future studies are needed to perform tests over a period of time to examine the potential adaptation effect to teleoperation under latency.

## 5. Conclusion

This study reported how the operators' neural processes were impacted by the teleoperation latency and a latency mitigation method through functional connectivity analysis with fNIRS data. Indicated by increased connectivity ratio and connectivity strengths, the results showed that the presence of latency in teleoperation increased functional connectivity within and between prefrontal and motor cortexes, representing higher levels of cognitive load, anxiety, and challenges in motion planning and control. While maintaining the same visual latency, providing real-time haptic feedback reduced the average functional connectivity in all cortical networks and showed a significantly different connectivity ratio within prefrontal and motor cortical networks. This implies that this latency-mitigation method impacts neural paths, especially those related to the cognitive, decision-making, and sensorimotor processes. The performance results (placement error and time on task) showed the worst performance in the all-delayed condition, which echoes the neural activity patterns. The findings from this study show that neurological evidence that latency in teleoperation is detrimental to operators mainly due to the increased level of cognitive barrier, and creating an optional sensorimotor process could potentially mitigate such an effect.

**Key points**:

- Latency between command, execution, and feedback in teleoperation can impair performance and affect operators' mental state. The neural underpinnings of these effects are not well understood.
- A human subject experiment (n = 41) of a simulated remote robot manipulation task was performed.
- Delayed haptic and visual feedback in teleoperation significantly reduced performance and increased functional connectivity in full connection, prefrontal, and motor cortex networks.
- Real-time haptic feedback could effectively reduce the average functional connectivity in all cortical networks and show a significantly different connectivity ratio within prefrontal and motor cortical networks.
- This research can inform the design of ergonomic teleoperation systems that mitigate the effects of latency.

**Biographies:**

Yang Ye: Research Assistant in the ICIC lab, University of Florida; BS (2019), Building Engineer, Hong Kong Polytechnic University; MS (2020), Computational Science, Imperial College London.

Tianyu Zhou: Post-Doc Researcher in the ICIC lab, University of Florida; BS (2017), Process Equipment and Control Engineering, Dalian University of Technology; MS (2019), Mechanical Engineering, University of Florida; PhD (2023) Civil Engineering, University of Florida.

Qi Zhu: UIUX Researcher, National Institute of Standards and Technology (NIST); BA (2014), Communication and Advertising, Huazhong University of Science and Technology; MS (2016), Computer Science, National Taiwan University; PhD (2022), Civil Engineering, University of Florida.

William Vann: Former Research Assistant in the ICIC lab, University of Florida; BS (2015), Agriculture & Biological Engineering, University of Florida; MS (2019), Mechanical Engineering, University of Florida. PhD (2023) Civil Engineering, University of Florida.

Eric Du: Professor, Engineering School of Sustainable Infrastructure & Environment, University of Florida. BS (2004), Civil Engineering, Tianjin University; MS (2007), Enterprise Management, Tianjin University, and PhD (2012), Construction, Michigan State University; Corresponding Author.

**Data Availability**

All data used in this paper can be found at: https://www.dropbox.com/scl/fo/4qcvc4v3nxhc9in0oi6bo/h?rlkey=15eds1e7on4vjkstannruok5r&dl=0